\documentclass[conference]{IEEEtran}
\usepackage{graphicx} 
\usepackage{times}
\usepackage{mathtools}
\usepackage{amsmath}
\usepackage{amssymb}
\usepackage{algorithm}      
\usepackage{algpseudocode}  

\usepackage{algorithm}
\usepackage{algorithmicx}
\usepackage{algpseudocode}

\algrenewcommand\algorithmicrequire{\textbf{Input:}}
\algrenewcommand\algorithmicensure{\textbf{Output:}}

\usepackage{caption}
\usepackage{url}
\usepackage{float} 
\usepackage{xcolor} 

\usepackage{geometry}
\date{}
\begin{document}

\title{Sequential BP-based Decoding of QLDPC Codes}

\author{
Mohsen~Moradi$^*$,
Salman~Habib$^\dag$,
Vahid~Nourozi$^*$,
and David~G.~M.~Mitchell$^*$\\
$^*$Klipsch School of Electrical and Computer Engineering, New Mexico State University\\ 
Email: \{moradi23, nourozi, dgmm\}@nmsu.edu, \\
$^\dag$Dept. of Electrical and Computer Engineering, Texas A \& M University-Texarkana \\
Email: shabib@tamut.edu
}

\maketitle
\begin{abstract}
Quantum low-density parity-check (QLDPC) codes are a leading approach to quantum error correction, yet conventional belief propagation (BP) decoders often perform poorly, primarily due to non-convergence {exacerbated by stabilizer constraints, which induce short cycles and degeneracy}. 
We propose two scheduling variants, sequential check node scheduling (SCNS) and sequential variable node scheduling (SVNS), that improve BP’s error-correction ability by processing check nodes (CNs) or variable nodes (VNs), respectively, in a fixed order, stabilizing message updates and reducing stalls. We also employ this technique to an improved BP-variant called BP guided decimation (BPGD), where symbols are progressively fixed during decoding iterations. Here, we demonstrate that the \emph{sequential BPGD} (SBPGD) decoder can further improve the convergence properties and performance of the decoder. On standard QLDPC benchmarks under a Pauli-\(X\) noise model, our sequential schedules are shown to lower the block error rate relative to conventional BP, and SBPGD outperforms BPGD while using significantly fewer decimation rounds, translating to lower computational cost. These results demonstrate that changing the update schedule, without altering the code, can improve both the reliability and efficiency of BP-based decoding for QLDPC codes. For the \(\bigl[\![1922,50,16]\!\bigr]\) C2 hypergraph-product code with independent \(X\) errors, SVNS-BP surpasses BP–OSD–0 in error correction at roughly the same complexity as standard BP.

\end{abstract}

\section{Introduction}
Low-density parity-check (LDPC) codes are foundational in modern error correction since their sparse Tanner graphs enable scalable, near-capacity belief-propagation (BP) decoding~\cite{gallager1962ldpc,richardson2008mct}.
In quantum error correction (QEC), recent construction breakthroughs have elevated quantum LDPC (QLDPC) codes to a leading candidate for fault tolerance. For example, 
hypergraph-product (HGP) codes provide explicit positive-rate families achieving distance scaling as $\Omega(\sqrt{n})$~\cite{Tillich} in the code length $n$, 
quantum expander codes support linear-time decoding under local noise~\cite{leverrier2015expander}, 
balanced-product codes improve the simultaneous scaling of code dimension and distance \cite{breuckmann2021balanced}, 
and non-abelian lifted products resolve the QLDPC conjecture by exhibiting asymptotically good families~\cite{panteleev2022good}. 
{More broadly, other stabilizer code families have also seen progress \cite{yang2025spatially, Vahid, google2025quantum}.}
However, \emph{decoding} remains as a practical bottleneck.

Directly applying BP to decode QLDPC codes will typically not result in good performance for two fundamental reasons.
First, stabilizer commutation enforces many \emph{short cycles}, which violates the independence assumptions underlying BP decoders.
Second, \emph{degeneracy} (that is, multiple Pauli error patterns sharing the same syndrome) creates symmetric pseudo-codewords that stall or misdirect BP’s fixed-point dynamics~\cite{poulin2008iterative}.
In practice, plain BP frequently oscillates, fails to converge, or converges to the wrong logical coset, especially at low physical error rates.

One promising strategy to combat this undesirable decoder behavior augments BP with \emph{post-processing}, notably ordered-statistics decoding (BP-OSD)~\cite{panteleev2021degenerate} and stabilizer inactivation (BP-SI)~\cite{du2022stabilizer}, which can close much of the performance gap at the cost of solving linear systems and additional heuristics. In contrast, belief-propagation-guided decimation (BPGD) \cite{yao2023bpgd} pursues an alternative approach: it interleaves short BP bursts with \emph{decimation} steps that freeze the most reliable variables, injecting asymmetry and shrinking the search space. 
BPGD substantially reduces non-convergence and can match the performance of BP-OSD/BP-SI without expensive post-processing. 
Follow-up work explores soft and soft-hard variants that integrate self-reinforcement into the updates and periodically introduce hard fixes, trading  complexity for further gains~\cite{alinia2025decimation}.

In classical BP decoding, sequential scheduling of check node (CN) and variable node (VN) updates is known to improve decoding performance. For example, one can select the CNs to be updated sequentially (in some prescribed order), where the BP decoder can be scheduled to update all the VNs only within the subgraph of each CN, then proceed to the next \cite{ zhang2005shuffled,sharon2007efficient,habib2021belief,hosseinzadeh2025layered}. Consequently, a VN may be updated multiple times within a single iteration. The advantage of this approach is that once a CN is updated, the newly generated messages to its neighboring VNs are used to refine the beliefs of subsequently scheduled CNs, potentially improving the accuracy of message propagation in the same iteration as well as enhancing algorithmic convergence. In contrast, in the flooding schedule, all CNs are updated simultaneously using the VN-to-CN messages computed in the previous iteration. For LDPC codes, optimizing the CN update order yields measurable gains~\cite{habib2021belief}. 
For polar codes, whose Tanner graphs have girth~4, a tailored sequential schedule can improve BP performance by more than $1$\,dB in some instances~\cite{moradi2025enhancing}. 
Both lines of work also cast schedule selection as a learning problem, employing reinforcement learning to discover effective update orders for LDPC~\cite{habib2021belief} and polar codes~\cite{moradi2025enhancing}.

In this paper, we propose to apply sequential scheduling for QLDPC codes to address the degeneracy and convergence challenge for BP. 
Our aim is to reduce \emph{non-convergence} at near-BP complexity. 
We formalize two sequential schedules for syndrome-domain QLDPC decoding: \emph{sequential VN scheduling} (SVNS) and \emph{sequential CN scheduling} (SCNS), which immediately reuse the freshest messages within each sweep. 
We further integrate these schedules into BPGD to obtain \emph{sequential BPGD} (SBPGD) with finite clamping and standard stopping. 
This combination reduces non-convergence and lowers the number of decimation rounds while also maintaining complexity comparable to flooding BP. Our numerical results show that SVNS-BP and SCNS-BP both yield improved error-correction performance (up to two or one order of magnitude, respectively) compared with flooding BP under the same cap on message-passing iterations, while reducing CN to VN message traffic by up to \(85\%\). In addition, the proposed SVNS-BPGD decoder achieves better error correction than BPGD with significantly fewer average decimations, indicating that sequential scheduling is a promising approach for improving BP-based decoders for QLDPC codes.

{Our method modifies only the schedule of the syndrome-processing decoder and does not increase the quantum memory footprint (the number of physical qubits $n$). By reducing message traffic and accelerating convergence, it can reduce decoding latency, which may indirectly relax coherence-time pressure in fault-tolerant operation. In terms of decoder memory, our method does not increase storage requirements beyond standard BP (it stores the same per-edge messages and node beliefs); it only changes the update order.}


\section{Background}\label{sec:background}
\subsection{Stabilizer Formalism}
An $[[n,k]]$ quantum stabilizer code uses $n$ physical qubits to reliably store and protect $k$ logical qubits from the effects of noise and error in a quantum system.
A pure state of a single qubit corresponds to a unit-norm vector in the complex Hilbert space $\mathbb{C}^2$.
The Pauli operators acting on a single qubit are represented by the  Hermitian matrices
$$X=\begin{bmatrix}0&1\\[2pt]1&0\end{bmatrix},
Z=\begin{bmatrix}1&0\\[2pt]0&-1\end{bmatrix},
Y=\begin{bmatrix}0&-i\\[2pt]i&0\end{bmatrix}, 
I=\begin{bmatrix}1&0\\[2pt]0&1\end{bmatrix},$$
where $i = \sqrt{-1}.$
For a system of $n$ qubits, the state space is given by the $n$-fold tensor product of the two-dimensional Hilbert space, denoted as $\mathbb{C}^{\otimes n}_2$.
The \(n\)-qubit Pauli group \(\mathcal{P}_n\) consists of \(n\)-fold tensor products of \(\{\pm 1,\pm i\}\!\cdot\!\{I,X,Y,Z\}\). 

For binary vectors \(a,b\in\mathbb{F}_2^n\), we define the \(n\)-fold Pauli operator
\[
P(a,b)\;=\;\bigotimes_{i=1}^n X^{a_i} Z^{b_i}.
\]
All operations in this paper are over \(\mathbb{F}_2\). 
 
A (binary) stabilizer code on $n$ physical qubits is specified by an abelian subgroup $\mathcal{S}\subset \mathcal{P}_n$ not containing $-I^{\otimes n}$ and defined as 
\begin{equation}
\mathcal{C} \;=\; \left\{\, |\psi\rangle \in (\mathbb{C}^2)^{\otimes n} \;:\; M\,|\psi\rangle = |\psi\rangle,\ \forall\, M \in \mathcal{S} \right\}.
\end{equation}
The code space is the $+1$ common eigenspace of $\mathcal{S}$ and has dimension $k$ when $\mathcal{S}$ has $n{-}k$ independent generators. The subgroup $\mathcal{S}$ is referred to as the \emph{stabilizer group}, and the Pauli operators in $\mathcal{S}$ are called the \emph{stabilizers}.
The code space consists of all states in $(\mathbb{C}^2)^{\otimes n}$ stabilized by $\mathcal{S}$.

In the symplectic representation, $\mathcal{S}$ is generated by the rows of the \emph{stabilizer matrix}
\begin{equation}
H \;=\; [\,H_X \; H_Z\,],\qquad H_X,H_Z\in\mathbb{F}_2^{m_s\times n},
\end{equation}
where each row $(h_x,h_z)$ corresponds to the stabilizer $P(h_x,h_z)$. Commutation of all stabilizers is equivalent to the quadratic constraint
\begin{equation}
H_X H_Z^\top \;+\; H_Z H_X^\top \;=\; 0. 
\end{equation}
The code distance is the minimum weight over $N(\mathcal{S})\setminus \mathcal{S}$, with $N(\mathcal{S})$ the normalizer of $\mathcal{S}$ in $\mathcal{P}_n$. 
We denote a quantum stabilizer code $\mathcal{C}$ with a minimum distance of $d$ as $[[n,k,d]]$.

As an important subclass of stabilizer codes, we focus on the \emph{Calderbank--Shor--Steane} (CSS) codes \cite{calderbank1996good, steane1996multiple}, in which each stabilizer operator involves either only Pauli-$X$ or only Pauli-$Z$ terms. In matrix form, a CSS code can be expressed as
\begin{equation}
H_X \;=\;
\begin{bmatrix}
0 \\[2pt]
G_2
\end{bmatrix},
\qquad
H_Z \;=\;
\begin{bmatrix}
H_1 \\[2pt]
0
\end{bmatrix},
\qquad
H \;=\;
\begin{bmatrix}
0 & H_1 \\[2pt]
G_2 & 0
\end{bmatrix},
\end{equation}
where $H_1 \in \mathbb{F}_2^{(n-k_1)\times n}$ denotes the parity-check matrix of a classical linear code $\mathcal{C}_1$ with parameters $[n, k_1]$, and $G_2 \in \mathbb{F}_2^{k_2\times n}$ represents the generator matrix of another classical code $\mathcal{C}_2$ with parameters $[n, k_2]$. 
The commutativity condition for the stabilizers
simplifies to
\begin{equation}
G_2 H_1^{\top} = 0,
\end{equation}
which is equivalent to the inclusion relation $\mathcal{C}_2 \subseteq \mathcal{C}_1$. 
Throughout this work, we adopt this CSS framework.

\subsection{Error Model}
For a stabilizer code $\mathcal{C}$, an encoded quantum state $|\psi\rangle$ may be affected by an $n$-qubit Pauli error 
$E = P(x, z) \in \mathcal{P}_n$, resulting in the corrupted state $E|\psi\rangle$. 
The objective of a decoder is to identify and correct such an error by performing measurements on all stabilizers specified by the matrix $H$. The outcomes of these stabilizer measurements form a binary \emph{syndrome vector} of length $m_s$, expressed as
\begin{equation}
s \;=\; (x, z)\,\Lambda\, H^{\top} = (xH_Z^{\top}, zH_X^{\top}) \;\in\; \mathbb{F}_2^{m_s},
\end{equation}
where $\Lambda$ is the symplectic matrix. 
This measured syndrome $s$ is then passed to the decoder for error estimation.

Based on the decoder’s output, the result of the decoding process can be classified into the following categories:
\begin{itemize}
    \item \textbf{Exact recovery:} the estimated error $\hat{E}$ matches the true error, i.e., $\hat{E} = E$;
    \item \textbf{Degenerate recovery:} the estimate differs from $E$ by a stabilizer, $\hat{E}E \in \mathcal{S}$;
    \item \textbf{Logical error:} the estimate differs from $E$ by a logical operator, $\hat{E}E \in N(\mathcal{S})\setminus \mathcal{S}$;
    \item \textbf{Failure:} the decoder cannot find an error consistent with the measured syndrome.
\end{itemize}

In this paper we focus on independent Pauli-$X$ error channel with probability $p_x$. The Pauli error has the form $E = P(X,0)$, where $X \in \{0,1\}^n$ is a binary random vector whose entries are independent with $\Pr[X_i=1]=p_x$ and $\Pr[X_i=0]=1-p_x$. For the CSS codes considered here, we only need the syndrome component
\[
s_x = X H_1^{\top},
\]
where $H_1$ is the submatrix of $H_z$ as explained above.

{Our contribution is schedule-level and does not rely on properties unique to the independent $X$-noise model; it primarily targets the convergence issues caused by short cycles and degeneracy. For CSS codes under depolarizing noise, one can decode the $X$ and $Z$ components via two coupled (or iterated) syndrome-domain BP decoders, and the same sequential schedules can be used for each component. For non-CSS codes, the same idea applies to quaternary/symplectic BP variants by adopting the same sequential order over nodes/edges while keeping the underlying update equations unchanged.
We leave studies under broader noise models and non-CSS families to future work.}

\subsection{BP for QLDPC}
BP is an iterative message-passing algorithm that operates on the Tanner graph induced by a code’s parity-check matrix. BP was first introduced for LDPC codes \cite{gallager1962ldpc} and also has been applied to quantum LDPC codes \cite{poulin2008iterative}.
We employ \emph{syndrome-domain} BP (sum–product) on the Tanner graph of the parity-check matrix $H_1$. 
The Tanner graph consists of VNs \(V=\{v_1, v_2, \ldots, v_n\}\), corresponding to the columns of \(H_1\), and CNs \(C=\{c_1, c_2, \ldots, c_m\}\), corresponding to the rows of \(H_1\). A VN \(v_i\) is connected to a CN \(c_j\) if and only if \(H_1(j,i)=1\).
For a VN $v$, $\partial v$ denotes the set of all CNs connected to $v$, and for a CN $c$, $\partial c$ denotes the set of all VNs connected to $c$.

For decoding, the channel LLRs are initialized as $\mu_v=\log\frac{1-p_x}{p_x}$ for each VN $v$. For CN $c$ and VN $v$, the standard sum-product updates are
\begin{align}
m_{c\to v} &= 2(-1)^{s(c)} \tanh^{-1}\!\!\Bigg(\prod_{u\in \partial c\setminus \{v\}} \tanh\!\big(m_{u\to c}/2\big)\Bigg), \\
m_{v\to c} &= \mu_v + \sum_{c'\in \partial v\setminus \{c\}} m_{c'\to v},
\end{align}
with \emph{bias} (approximate marginal LLR) $m_v=\mu_v+\sum_{c\in\partial v} m_{c\to v}$ and hard decision 
$\hat{x}_v=1$ if $m_v<0$, and $\hat{x}_v=0$ otherwise.
The message-passing algorithm runs for at most a predefined number of iterations \(T\), with early stopping if the tentative estimate \(\hat{x}\) satisfies \(\hat{x} H^{\top} = s_x\).


\subsection{Belief Propagation Guided Decimation (BPGD)}
\label{sec:bpgd}
BPGD decoding  combines short bursts of BP with \emph{guided decimation} that \emph{freezes} highly reliable variables \cite{ yao2023bpgd,alinia2025decimation}. After $T$ BP iterations, the biases $m_v$ are computed; if the tentative decision $\hat x$ fails the syndrome test, select $i=\arg\max_v |m_v|$ and \emph{clamp} (decimate) its channel LLR to a large \emph{finite} value, then continue BP on the modified graph. This simple asymmetry injection progressively shrinks the search space and breaks degeneracy-induced traps.
BPGD (both binary and GF(4)) \cite{alinia2025decimation, yao2023bpgd} substantially reduces non-convergence and often reaches BP–OSD/BP–SI performance without solving linear systems. In these papers, they also report that most residual BPGD errors are still non-convergences, not logical mistakes, underscoring the value of decimation.

\section{Sequential Scheduling}\label{sec:sequential}

In this section, we present our proposed sequential VN scheduling (SVNS) and sequential CN scheduling (SCNS) schemes. 

\subsection{Sequential CN Scheduling (SCNS)}

\begin{algorithm}[t]
\small
\caption{Sequential Check Node Schedule (SCNS) BP Decoding}
\label{alg:rsbp-cn}
\begin{algorithmic}[1]
\Require $H_1\in\{0,1\}^{m\times n}$ with neighborhoods $\partial c,\partial v$; syndrome $s_x\in\{0,1\}^m$; $p_x$; channel LLRs $\mu_v=\log\!\frac{1-p_x}{p_x}$; fixed CN order $\pi_C$; iteration cap $T$
\Ensure $\hat{x}\in\{0,1\}^n$, convergence flag, iterations used
\State \textbf{Initialization:} $\text{bitLLR}_v \gets \mu_v$; $m_{v\to c}\gets \mu_v$ for all edges; $\sigma_c\gets (-1)^{s_x(c)},\ \forall c$
\For{$t=1$ to $T$}
  \ForAll{$c \in \pi_C$}
    \State $E_c \gets \{(v,c): v\in\partial c\}$
    \ForAll{$(v,c)\in E_c$}
      \State $P \gets \sigma_c \prod\limits_{u\in\partial c\setminus \{v\}}\tanh\!\big(\tfrac{1}{2}m_{u\to c}\big)$
      \State $m_{c\to v} \gets 2\,\mathrm{atanh}(P)$
    \EndFor
    \ForAll{$(v,c)\in E_c$}
      \State $S \gets \mu_v + m_{c\to v}$
      \ForAll{$c'\in\partial v\setminus \{c\}$}
        \State $Q \gets \sigma_{c'} \prod\limits_{u\in\partial c'\setminus \{v\}}\tanh\!\big(\tfrac{1}{2}m_{u\to c'}\big)$
        \State $S \gets S + 2\,\mathrm{atanh}(Q)$
      \EndFor
      \State $\text{bitLLR}_v \gets S$
      \State $m_{v\to c} \gets \text{bitLLR}_v - m_{c\to v}$
    \EndFor
  \EndFor
  \State $\hat{x}_v \gets \mathbb{I}[\text{bitLLR}_v<0]$, $\forall v$
  \State $\hat{s}_x(c) \gets \big(\sum_{v\in\partial c}\hat{x}_v\big)\bmod 2$,  $\forall c$
  \If{$\hat{s}_x=s_x$} \Return $(\hat{x},\texttt{true},t)$ \EndIf
\EndFor
\State \Return $(\hat{x},\texttt{false},T)$
\end{algorithmic}
\end{algorithm}

We present the SCNS procedure in Algorithm~\ref{alg:rsbp-cn}.
For a fixed iteration cap \(T\) and a fixed ordering (randomly selected in this paper) of CNs \(\pi_C\), the decoder visits CNs sequentially within each iteration.
For the current CN \(c\), lines 5--8 propagate messages from the CN to all neighboring VNs. Next, lines 9--17 update the VNs connected to \(c\) and send messages back to this CN. During this step, each affected VN also needs to refresh the messages from its other incident CNs before returning its update. 
After each full sweep, the decoder forms hard decisions, verifies the measured syndrome, and terminates early if they match, reporting the iteration count. If no match is found within \(T\) iterations, the decoder reports non-convergence. 
All other details follow the standard syndrome-domain BP specified in Section~\ref{sec:background}.

\subsection{Sequential VN Scheduling (SVNS)}

\begin{algorithm}[t]
\small
\caption{Sequential Variable Node Schedule (SVNS) BP Decoding}
\label{alg:rsbp-vn}
\begin{algorithmic}[1]
\Require $H_1\!\in\!\{0,1\}^{m\times n}$ with neighborhoods $\partial c,\partial v$; measured $X$-syndrome $s_x\!\in\!\{0,1\}^m$;  $p_x$; channel LLRs $\mu_v=\log\!\frac{1-p_x}{p_x}$; fixed VN order $\pi_V$; iteration cap $T$
\Ensure Hard estimate $\hat{x}\in\{0,1\}^n$, convergence flag, iterations used
\State \textbf{Initialization:} $\text{bitLLR}_v \gets \mu_v,\ \forall v$;\quad $m_{v\to c}\gets \mu_v,\ \forall (v,c)\text{ with }v\in\partial c$;\quad $\sigma_c\gets (-1)^{s_x(c)},\ \forall c$
\For{$t=1$ to $T$}
  \ForAll{$v \in \pi_V$} 
    \State $E_v \gets \{(v,c): c\in\partial v\}$ 
    \ForAll{$(v,c)\in E_v$} 
      \State $P \gets \sigma_c\!\!\!\prod\limits_{u\in\partial c\setminus v}\!\!\! \tanh\!\big(\tfrac{1}{2}m_{u\to c}\big)$
      \State $m_{c\to v} \gets 2\,\mathrm{atanh}(P)$
    \EndFor
    \State $\text{bitLLR}_v \gets \mu_v + \sum_{c\in\partial v} m_{c\to v}$ 
    \ForAll{$(v,c)\in E_v$}
      \State $m_{v\to c} \gets \text{bitLLR}_v - m_{c\to v}$ 
    \EndFor
  \EndFor
  \State $\hat{x}_v \gets \mathbb{I}[\text{bitLLR}_v<0],\ \forall v$
  \State $\hat{s}_x(c) \gets \big(\sum_{v\in\partial c}\hat{x}_v\big)\bmod 2,\ \forall c$
  \If{$\hat{s}_x=s_x$} \Return $(\hat{x},\texttt{true},t)$ \EndIf
\EndFor
\State \Return $(\hat{x},\texttt{false},T)$
\end{algorithmic}
\end{algorithm}

We also present our proposed SVNS procedure in Algorithm~\ref{alg:rsbp-vn}. 
For a fixed iteration cap \(T\) and a fixed ordering of the VNs (also chosen to be random in this paper) \(\pi_V\), the decoder visits VNs sequentially within each iteration. 
For the current VN \(v\), lines 5–8 first gather fresh CN to VN messages from all incident CNs. 
Line 9 updates the VN belief using these messages and the channel prior. 
Lines 10–12 then send updated VN to CN messages back along every edge incident to \(v\). 
After each full sweep over \(\pi_V\), the decoder forms hard decisions, verifies the measured syndrome, and terminates early if it matches; otherwise it proceeds to the next iteration. 
If no match is found within \(T\) iterations, the decoder reports non-convergence. 
Similarly, all other details follow the standard syndrome-domain BP specified in Section~\ref{sec:background}.

\subsection{Sequential BPGD}

Sequential BPGD replaces the flooding BP subroutine inside BPGD with SVNS or SCNS. Each \emph{round} runs $T$ SVNS/SCNS iterations, checks the syndrome, and if unsatisfied, decimates the most reliable VN (finite clamping) before the next round. Because sequential schedules typically reach higher reliabilities in fewer message updates, one expects to observe fewer decimation rounds, especially for codes where flooding BP exhibits frequent stalls.

\section{Numerical Results}\label{sec:experiments}
In this section, we report results of a software-based implementation of the BP and BPGD decoders with and without the sequential schedules as proposed in Section \ref{sec:sequential}.

\subsection{Sequential CN Scheduling BP (SCNS-BP)}

\begin{figure}[t]
  \centering
  \includegraphics[width=\linewidth]{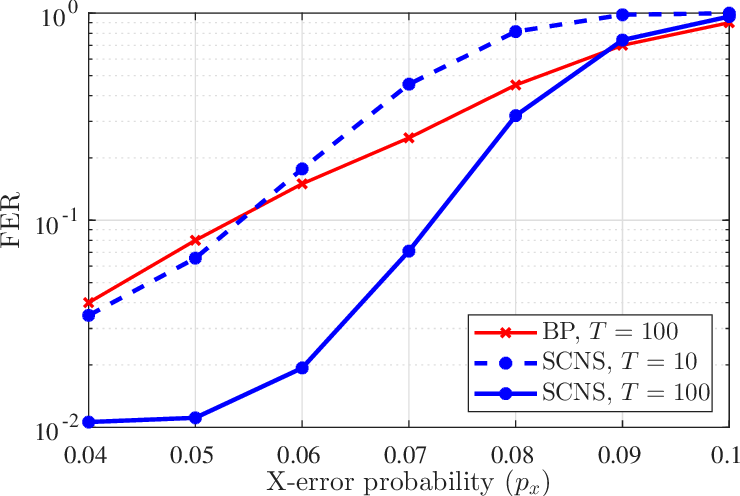}
  \caption{HGP code \(C2\) \([[1922,50,16]]\) with flooding BP vs. our proposed SCNS-BP decoding algorithm.}
  \label{fig:C2_SCNS}\vspace{-2mm}
\end{figure}

\begin{figure}[t] 
  \centering
  \includegraphics[width=1\linewidth]{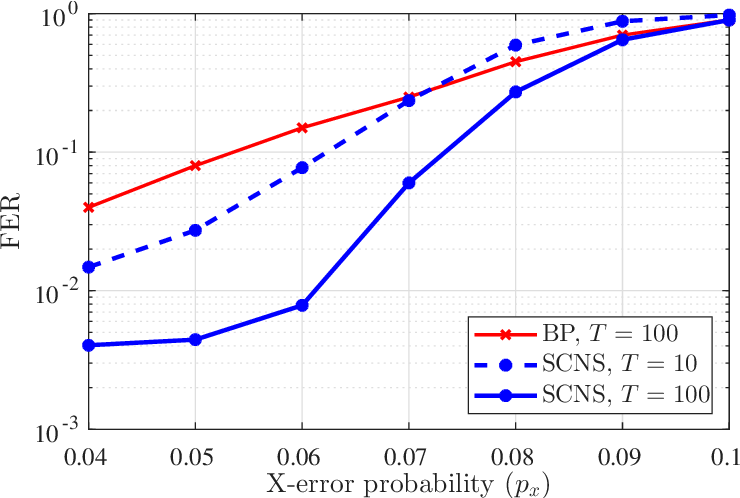}
  \caption{Code \(B1\) $[[882, 24, 18\leq d \leq 24]]$ with flooding BP vs. our proposed SCNS-BP decoding algorithm.}\vspace{-2mm}
  \label{fig:B1_SCNS}
\end{figure}

In Fig.~\ref{fig:C2_SCNS}, we compare SCNS with standard flooding BP on the \([[1922,50,16]]\) HGP code \(C2\) \cite{panteleev2021degenerate} across different iterations of $T=100$ and $T=10$. We observe that a sequential schedule (randomly selected, not optimized) is seen to offer an advantage over the flooding BP decoder. For the same number of global iterations, SCNS-BP achieves a lower frame error rate (FER), and with fewer iterations it matches flooding BP at low $p_x$ values.\footnote{We note that the same number of global iterations will have an equal number of CN updates between BP and SCNS, however the SCNS will have an increased number of VN updates (a complexity discussion is included below).} We observe the emergence of an error floor for $p_x \leq 0.05$.  Fig.~\ref{fig:B1_SCNS}, shows a similar experiment for the \([[882, 24, 18\leq d \leq 24]]\) code \(B1\) \cite{panteleev2021degenerate} across different iterations of $T=100$ and $T=10$. 
Similar conclusions are drawn in this case. For the same number of global iterations, SCNS-BP achieves an order-of-magnitude lower FER for $p_x\leq 0.06$ with an error floor emerging at $p_x\leq 0.05$. Improved performance is again observed with fewer global iterations ($T=10$) at low $p_x$ values.

\subsection{Sequential VN Scheduling BP (SVNS-BP)}

\begin{figure}[t] 
  \centering
  \includegraphics[width=1\linewidth]{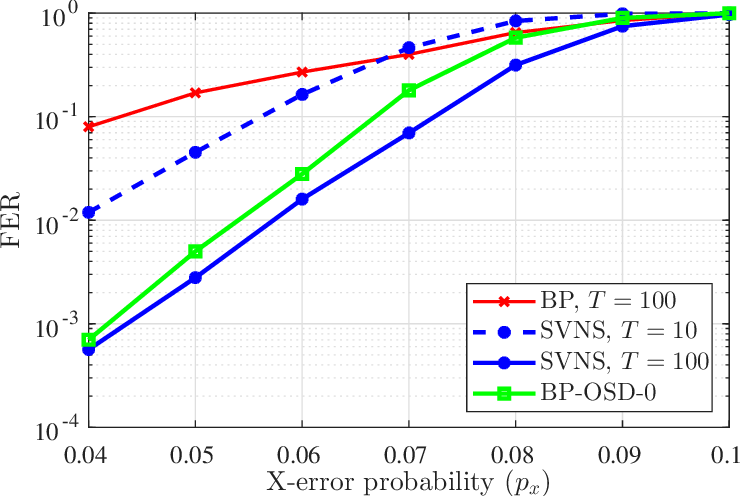}
  \caption{HGP code \(C2\) \([[1922,50,16]]\) with flooding BP vs. our proposed SVNS-BP decoding algorithm.}
  \label{fig:C2_SVNS}
\end{figure}

In Fig.~\ref{fig:C2_SVNS}, we compare our proposed SVNS BP decoder versus standard flooding BP on the \(\bigl[\![1922,50,16]\!\bigr]\) HGP code \(C2\) across different iteration counts. We observe significant improvement when compared to flooding decoding. For the same number of global iterations, SVNS achieves approximately two-orders-of-magnitude lower FER for small $p_x\leq 0.05$. Even with fewer global iterations ($T=10$), it surpasses flooding BP by up to one order of magnitude at \(p_x = 0.04\). 
In this figure, we also compare against BP-OSD-0 with \(T=100\) iterations, which uses post-processing and requires solving a linear system of equations. We see SVNS can outperform BP-OSD-0 without the need for post-processing.

\begin{figure}[t] 
  \centering
  \includegraphics[width=1\linewidth]{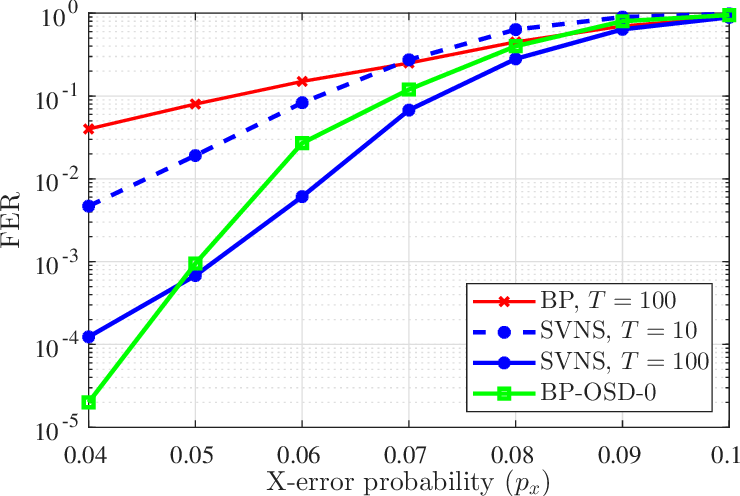}
  \caption{Code \(B1\) $[[882, 24, 18\leq d \leq 24]]$ with flooding BP vs. our proposed SVNS-BP decoding algorithm.}
  \label{fig:B1SVNS}
\end{figure}

Fig.~\ref{fig:B1SVNS} shows the same experiment for the \(\bigl[\![882,\,24,\,18\le d \le 24]\!\bigr]\) code \(B1\) under iteration budgets \(T\in\{10,100\}\).
Similar conclusions are drawn, where for the same number of global iterations \(T=100\), SVNS\textendash BP attains more than two orders of magnitude lower FER for small $p_x$; with fewer iterations \(T=10\), it still surpasses flooding BP by up to one order of magnitude at low \(p_x\).
For reference, the figure also includes BP\textendash OSD\textendash 0 with \(T=100\), which uses post\textendash processing and requires solving a linear system. Here, we note that our SVNS was not able to outperform BP-OSD-0 at $p_x=0.04$, which is consistent with other BP-variant simulations (see, e.g., \cite{alinia2025decimation}).

\subsection{Message Complexity}
\begin{table*}[t]
\centering
\caption{Average Count of CN to VN Messages Propagated in Different Decoding Schemes for the decoders of the $\bigl[\!\bigl[1922,50,16\bigr]\!\bigr]$ C2 code with $T=100$ iterations.}
\label{tab:message_countsC2}
\begin{tabular}{|c||c|c|c|c|c|c|c|}
\hline
$p_x$ & 0.04 & 0.05 & 0.06 & 0.07 & 0.08 & 0.09 & 0.10 \\ \hline\hline
BP, CN to VN Messages           & 162376 & 221207 & 290289 & 360611 & 451905 & 543244 & 573832 \\ \hline
SCNS-BP, CN to VN Messages        & 28331  & 37073  & 56329  & 108660 & 261678 & 473746 & 565005 \\ \hline
SVNS-BP, CN to VN Messages        & 24272  & 33817  & 53736  & 106198 & 264007 & 476946 & 565575 \\ \hline
\end{tabular}
\end{table*}

Table~\ref{tab:message_countsC2} reports the average number of CN to VN messages per decoding attempt for flooding BP versus our SCNS and SVNS decoders for the \(\bigl[\![1922,50,16]\!\bigr]\) code \(C2\) with \(T=100\). At high error probabilities (\(p_x\in\{0.09,0.10\}\)), all schemes propagate a similar number of messages, reflecting widespread unsatisfied CNs that limit early termination. In the low-error regime (\(p_x\le 0.05\)), the sequential schedules reduce message traffic by more than \( 85\%\) relative to flooding BP (e.g., \(1.62\times10^5\) vs. \(2.43\times10^4\) CN to VN messages at \(p_x=0.04\)), since the sequential scheduling accelerates convergence. We note that SCNS and SVNS track each other closely across \(p_x\), with SVNS slightly lower at small \(p_x\). Overall, the reduced message propagation may translate to implementations with lower runtime without sacrificing FER.

\begin{table*}[t]
\centering
\caption{Average Count of CN to VN Messages Propagated in Different Decoding Schemes for the decoders of the $\bigl[\!\bigl[882,24,18\leq d \leq 24\bigr]\!\bigr]$ B1 code with $T=100$ iterations.}
\label{tab:message_counts_B1}
\begin{tabular}{|c||c|c|c|c|c|c|c|}
\hline
$p_x$ & 0.04 & 0.05 & 0.06 & 0.07 & 0.08 & 0.09 & 0.10 \\ \hline\hline
BP, CN to VN Messages           & 42153 & 64334 & 87911 & 111516 & 161806 & 216017 & 252394 \\ \hline
SCNS-BP, CN to VN Messages        & 8975  & 12050  & 17985  & 38810  & 100553 & 189483 & 244792 \\ \hline
SVNS-BP, CN to VN Messages        & 9011 & 12055  & 17901  & 36649  & 102773  & 190734 & 244239 \\ \hline
\end{tabular}\vspace{-3mm}
\end{table*}

Table~\ref{tab:message_counts_B1} presents the average CN to VN messages recorded per simulation for the \(\bigl[\![882,24,18\le d \le 24]\!\bigr]\) code \(B1\) at \(T=100\). 
In the low-error regime, both sequential schedules are seen to significantly reduce message traffic relative to flooding BP, for example, at \(p_x=0.04\) the counts drop from \(4.22\times10^4\) (BP) to \(\approx 9.0\times10^3\) (SCNS/SVNS), a \(\sim\!79\%\) reduction; at \(p_x=0.05\) the reduction is \(\sim\!81\%\). 
As \(p_x\) increases, the gap narrows and the schemes converge to similar counts. Again, both SVNS and SCNS track closely in terms of propagated messages across \(p_x\).

{Both schedules have the same asymptotic per-iteration cost order as BP (message passing on the Tanner graph), but as the tables show they can substantially reduce average message traffic and runtime via faster convergence/earlier stopping. We emphasize that degeneracy can produce symmetric pseudo-codewords that promote oscillation under flooding updates; sequential schedules break this synchrony by reusing the newest messages, which empirically reduces stalls and non-convergence.}

\subsection{Sequential BPGD (SBPGD)}

\begin{figure}[t] 
  \centering
  \includegraphics[width=1\linewidth]{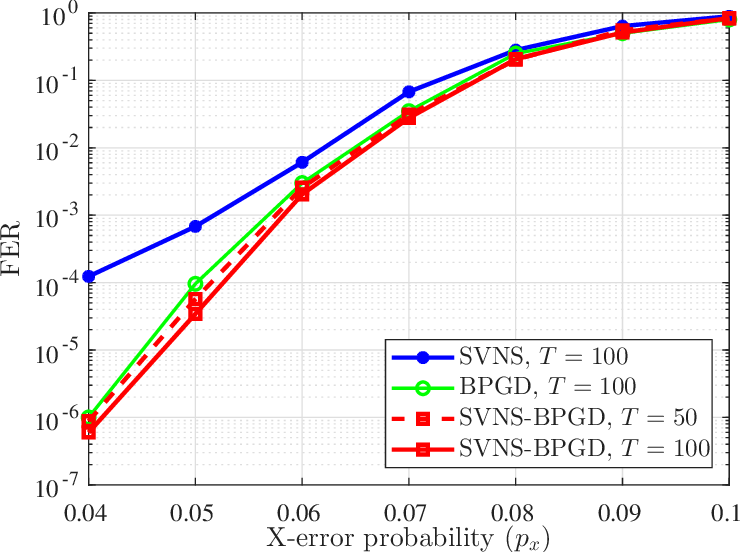}
  \caption{Code \(B1\)  with BPGD vs. our proposed SVNS-BPGD decoding algorithm with different number of BP iterations.}\vspace{-0mm}
  \label{fig:SVNC-BPGD}
\end{figure}

In Fig.~\ref{fig:SVNC-BPGD}, we report the performance of our SBPGD decoder for the \(\bigl[\![882,\,24,\,18\le d \le 24]\!\bigr]\) code \(B1\) with iteration caps \(T\!=\!50\) and \(T\!=\!100\), using SVNS for the scheduling. The results show that, even with fewer BP iterations, the sequentially aided BPGD slightly surpasses standard BPGD. For reference, we also include the SVNS curve for comparison. The main gain from this approach is seen in terms of the complexity. Table~\ref{tab:avg_decimations_b1} shows the average number of decimations for BPGD and our proposed SVNS-BPGD decoder, corresponding to Fig.~\ref{fig:SVNC-BPGD}.
As the results indicate, the sequential decoder achieves slightly better error-correction performance than BPGD while using a significantly smaller average number of decimations across all values of \(p_x\).

\begin{table}[t]
\centering
\caption{Average number of decimation steps per decoding attempt for BPGD and SVNS–BPGD on code \(B1\) across crossover probabilities \(p_x\) corresponding to Fig. \ref{fig:SVNC-BPGD}.}
\label{tab:avg_decimations_b1}
\begin{tabular}{|l||c|c|c|c|}
\hline
\multicolumn{1}{|c||}{Method} & \multicolumn{4}{c|}{\(p_x\)} \\ \cline{2-5}
\multicolumn{1}{|c||}{} & 0.04 & 0.05 & 0.06 & 0.07  \\ \hline\hline
BPGD, \(T=100\)      & 1.1469 & 1.4948 & 4.5460 & 41.6130  \\ \hline
SVNS–BPGD, \(T=100\) & 0.0056 & 0.0587 & 1.6657 & 28.0800  \\ \hline
SVNS–BPGD, \(T=50\)  & 0.0058 & 0.0762 & 2.1195 & 33.5910  \\ \hline
\end{tabular}\vspace{-1mm}
\end{table}

\section{Conclusions}\label{sec:conclusion}
We introduced two sequential scheduling algorithms for BP decoding of QLDPC codes: sequential CN scheduling (SCNS) and sequential VN scheduling (SVNS). In several regimes these schedules improve FER by more than two orders of magnitude compared with flooding BP. 
At low error probabilities, they also cut CN to VN message traffic by up to 85\%, indicating meaningful savings in computational effort. 
Embedding our sequential schedules within BPGD further improves both error correction and complexity relative to standard BPGD with comparable or fewer iterations. 
Altogether, the results suggest that sequential scheduling is a promising direction for decoding QLDPC codes. 
Future work includes optimizing the schedule (for example, via learning or adaptive policies), combining the schedules with other techniques beyond guided decimation, and extending the study to broader noise models or code families.

\section{Acknowledgment}
This material is based upon work supported by the National Science Foundation under Grant No. CCF-2145917.


\end{document}